\documentclass[twocolumn,pra,aps,superscriptaddress,showpacs]{revtex4}
\usepackage[latin9]{inputenc}
\usepackage{bm}
\usepackage{amsmath}
\usepackage{amssymb}
\usepackage{graphicx}
\usepackage{esint}
\usepackage{bm}
\usepackage{CJK}

\begin{document}

\title{Quantifying non-Markovianity for a chromophore-qubit pair in a super-Ohmic bath}

\author{Jing Liu}

\affiliation{Division of Materials Science, Nanyang Technological University,
50 Nanyang Avenue, Singapore 639798, Singapore}

\affiliation{Zhejiang Institute of Modern Physics, Department of Physics, Zhejiang
University, Hangzhou 310027, China}

\author{Kewei Sun}

\affiliation{Division of Materials Science, Nanyang Technological University,
50 Nanyang Avenue, Singapore 639798, Singapore}

\affiliation{School of Science, Hangzhou Dianzi University, Hangzhou 310018, China}

\author{Xiaoguang Wang}

\affiliation{Zhejiang Institute of Modern Physics, Department of Physics, Zhejiang
University, Hangzhou 310027, China}

\author{Yang Zhao}

\email{yzhao@ntu.edu.sg}

\affiliation{Division of Materials Science, Nanyang Technological University,
50 Nanyang Avenue, Singapore 639798, Singapore}

\begin{abstract}
An approach based on a non-Markovian time-convolutionless polaron
master equation is used to probe the quantum dynamics of a chromophore-qubit
in a super-Ohmic bath. Utilizing a measure of non-Markovianity based on dynamical fixed
points, we study the effects
of the environment temperature and the coupling strength on
the non-Markovian behavior of the chromophore in a super-Ohmic bath. It is found that an
increase in the temperature results in a reduction in
the backflow information from the environment
to the chromophore, and therefore, a suppression of
non-Markovianity. In the weak coupling regime,
increasing coupling strength will enhance the non-Markovianity,
while the effect is reversed in the strong coupling regime.
\end{abstract}


\maketitle

\section{Introduction}

Recent advances in spectroscopic techniques have allowed increasing deployments
of nonlinear optical measurements to probe dynamic properties of various condensed-matter and biological systems~\cite{Engel07,Lee07,Collini,Panit10,Sarovar10,Dijkstra,Intro,Intro2,Intro3,experiment,Brown,Reilly}.
An interesting example from
two-dimensional electronic spectroscopy studies is the conjecture that long-lasting quantum coherence may exist
photosynthetic light harvesting systems~\cite{Engel07,Lee07,Collini,Panit10,Sarovar10}. Nonlinear single molecule
spectroscopic techniques, such as hole-burning and three-pulse photon echo spectroscopy~\cite{experiment}, capable to generate truly homogeneous lineshapes by eliminating inhomogeneous broadening, have been used
to probe chromophores embedded in organic glasses, revealing a wide range of
spectral behaviors, where the coupling of chromophores to the surrounding
medium (solvent, glass, host crystal, protein, etc) may give rise to non-Markovian dynamics.

For low-temperature glasses and amorphous solids characterized by structural disorder, often only the two lowest energy levels of the double minimum potential need to be considered. Therefore, they can be modeled as a collection of two-level systems, and indeed such a model has been successfully employed to study their anomalous specific heat and thermal conductivity~\cite{Zeller, Heuer}.
With the local environment modeled as a
collection of flipping qubits which modulate the chromophore transition frequency~\cite{glass}, Suarez and Silbey~\cite{Suarez}
proposed a microscopic Hamiltonian to study the dynamics of a single
chromophore in glasses, and demonstrated its correlation with the stochastic sudden jump model~\cite{anderson}. Their
dressed microscopic Hamiltonian is taken as our starting point to investigate the dynamics of a central
chromophore embedded in a bath of qubits commonly found in low-temperature glasses.

In the aforementioned chromophore-qubit pair, the energy scales
for the vibronic relaxation and spin-phonon coupling are comparable placing the system-bath interaction
outside the usual weak coupling regime which is inaccessible to the traditional second-order
perturbation methods~\cite{Jang,Cheng,Kolli11}. Therefore,
non-perturbative approaches~\cite{Hughes,Prior,Thorwart,Nalbach},
including the numerically exact iterative path integral methods~\cite{Sahrapour},
sophisticated stochastic treatments of the system-bath models~\cite{Roden},
and hierarchical equation of motion approach~\cite{Ishizaki1}, have
subsequently been proposed to treat such systems of intermediate coupling. However,
these computationally intensive methods is inadequate to deal with
large systems or multiple-excitations.
Recently, the non-Markovian time-convolutionless polaron master equation
has been employed to describe the excitation dynamics in the multichromophoric
systems~\cite{Jang1,Jang2,Nazir,McCutcheon,Wu13,Kolli11}. The advantage
of this master equation is that it is capable of depicting the dynamics
in intermediate coupling regimes, handling initial non-equilibrium
bath states, as well as the spatially correlated environments.
This method has been successfully applied to study the dynamics of two coupled pseudo-spins in contact with a dissipative
bath and in addition, it was used to investigate the energy transfer of an extended spin-boson model by
including an additional spin bath~\cite{Wu13} .

Open quantum systems may exhibit interesting non-Markovian
features that have been drawing sustained attention~\cite{nM_intro,nM_intro2,rivas_review,nM_exper}.
How to quantify this behavior is one of the central
issues. Among various definitions of non-Markovianity that emerged \cite{Breuer09,Rivas10,Liu13,Lu10,Luo12,Chruscinski10},
one of the earliest, widely-used definitions was proposed by Breuer,
Laine and Piilo (BLP)~\cite{Breuer09}, which is based on the decreasing
monotonicity of the trace distance under the completely positive and trace-preserving
operations. One intuitive physical interpretation of this monotonicity
is that the information of distinguishability always flows from the
system to reservoir in a Markovian process. For a non-Markovian
dynamics, this monotonicity can be violated and the trace distance
may increase during the dynamics, indicating that the information
of distinguishability may flow back from the reservoir to system.

In this paper, we propose a new measure of non-Markovianity based
on the aforementioned mechanism for systems with dynamical fixed points. If $N$
is the number of the initial states one take for numerical calculation,
then this measure has a $\mathcal{O}(N)$ numerical advantage compared
with the BLP measure. By solving the time-convolutionless polaron
master equation of the chromophore-qubit pair, we find a
fixed point for the chromophore dynamics. Utilizing the non-Markovianity measure based on this fixed point,
we are allowed to quantify the non-Markovian behavior of the central chromophore.
Furthermore, we analyze the effects of the temperature and the coupling
between the chromophore and the qubit on the non-Markovianity,
and it is found that the temperature can suppress the backflow of
the information in our model. With respect to the effect of the chromophore-qubit coupling,
the situation is more complicated with differing influences of the coupling on the non-Markovianity in different regimes.
We will show that in the weak coupling regime,
the increase of the coupling strength can enhance the non-Markovianity, while
in the strong coupling regime, it will suppress the non-Markovianity.
In addition,  the non-Markovian behaviors of the chromophore-qubit pair and the corresponding
quasi-particle after the polaron transformation are also investigated. It is found that
the non-Markovian behavior vanishes after the polaron transformation
due to the much reduced coupling between the dressed particle and the
bath.

The paper is organized as follows. In Sec.~\ref{sec:MandD}, we introduce
the model and the time-convolutionless polaron master equation
of the chromophore-qubit pair with additional discussion on dynamical
fixed points of the chromophore. In Sec.~\ref{sec:NM}, we revisit
the BLP non-Markovianity measure and propose a new measure based on
the dynamical fixed points of the system. In Sec.~\ref{sec:Discussion},
we apply the new measure to our model and discuss the non-Markovian
behavior of the chromophore. Section~\ref{sec:Conclusion} draws
the conclusion of this work.

\section{Model and dynamics} \label{sec:MandD}

Chromophore is a term that commonly refers to a certain moiety of a large organic molecule that gives rise to its optical absorption and fluorescence properties, such as the pi-conjugated double bonds between carbon atoms in carotenoids, or the chlorine-type macrocyclic ring complexed by magnesium in chlorophylls. In the context of our work, this term refers to an entire molecule when it is embedded in a host environment, such as a pigment molecule in crystalline (e.g., pentacene in p-terphenyl crystal) or amorphous materials (perylene in polyethylene). In either context, a chromophore
can be simply modeled as a system with two electronic levels (the ground the excited state) whose transition frequency can be modulated due to its interaction with the host environment.

\begin{figure}
\includegraphics[width=7cm]{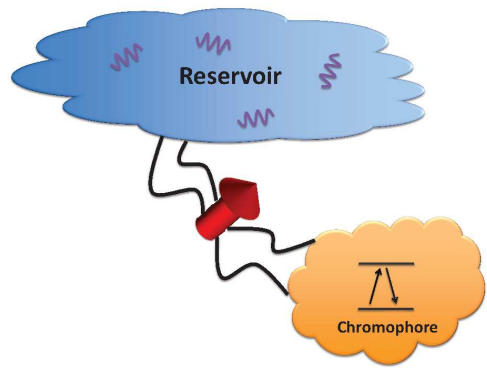}\caption{\label{fig:Schematics} Schematics of the model. The
chromophore interacts with the reservoir indirectly via a probe qubit. }
\end{figure}

In this paper, we consider a two-level chromophore coupled to a phonon bath
via a probe qubit, as shown in Fig.~\ref{fig:Schematics}. The Hamiltonian
can be written as~\cite{Suarez,Brown,Chen08}
\begin{eqnarray}
H & = & \frac{\omega_{0}}{2}\sigma_{z0}+\frac{\epsilon}{2}\sigma_{z1}-\frac{\Delta}{2}\sigma_{x1}
+\frac{a}{2}\sigma_{z0}\sigma_{z1}\nonumber \\
&  & +\sum_{k}\omega_{k}b_{k}^{\dagger}b_{k}+\sum_{k}g_{k}\left(b_{k}^{\dagger}+b_{k}\right)\sigma_{z1}.\label{eq:H}
\end{eqnarray}
Here $\sigma_{i0}:=\sigma_{i}\otimes \openone$ and $\sigma_{i1}:=\openone\otimes \sigma_{i}$ for $i=x,y,z$ and
$\sigma_{i}$ is a Pauli matrix. The subscript $0$ ($1$) represents the subspaces of chromophore (qubit).
We have set $\hbar=1$ in this Hamiltonian. $\omega_{0}$ and $\epsilon$
are the transition frequencies. $\Delta$ is the tunneling
matrix element. The coupling strength between the chromophore and qubit
is represented by the coefficient $a$, which is related
to the distance between the chromophore
and qubit, as well as the spin orientation of the qubit. $\sum_{k}\omega_{k}b_{k}^{\dagger}b_{k}$
is the Hamiltonian of the phonon bath, with $b_{k}^{\dagger}$, $b_{k}$
the creation and annihilation operators, respectively. $g_{k}$ represents
the coupling between the qubit and the reservoir. The Hamiltonian (\ref{eq:H})
can be also obtained through a unitary transformation from a Hamiltonian
in which the chromophore and the qubit are not directly coupled,
but both interact with a common reservoir~\cite{Brown,Suarez}.

%------------Figure 1--------------------
%-------------------------------------------

To facilitate dynamics calculation, we first perform a polaron transformation
on the Hamiltonian (\ref{eq:H}). The generator of this transformation reads
\begin{equation}
U=\frac{1}{2}B\sigma_{z1},
\end{equation}
where $B=\sum_{k}(g_{k}/\omega_{k})(b_{k}^{\dagger}-b_{k})$.
It is easy to verify the identify that $\exp(U)=\cosh(B/2)+\sigma_{z1}\sinh(B/2)$.
After the polaron transformation, the new Hamiltonian can be written as
\begin{equation}
\tilde{H}=e^{U}He^{-U}=\tilde{H}_{0}+\tilde{H}_{\mathrm{I}}+H_{\mathrm{b}}.
\end{equation}
Here the effective quasi-particle Hamiltonian $\tilde{H}_{0}$ has the form
\begin{equation}
\tilde{H}_{0}=\frac{\omega_{0}}{2}\sigma_{z0}+\frac{\epsilon}{2}\sigma_{z1}-\frac{\Delta}{2}\sigma_{x1}\Theta+\frac{a}{2}\sigma_{z0}\sigma_{z1}-\sum_{k}\frac{g_{k}^{2}}{4\omega_{k}}.\label{eq:H_0}
\end{equation}
The bath Hamiltonian is unchanged as $H_{\mathrm{b}}=\sum_{k}\omega_{k}b_{k}^{\dagger}b_{k}$
and the effective interaction Hamiltonian $\tilde{H}_{\mathrm{I}}$ can be
expressed by
\begin{equation}
\tilde{H}_{\mathrm{I}}=-\frac{\Delta}{2}\left[\sigma_{x1}\left(\cosh B-\Theta\right)+i\sigma_{y1}\sinh B\right], \label{eq:inter_Hamil}
\end{equation}
where $\Theta=\langle\cosh B\rangle=\langle\exp B\rangle$, and $\langle\cdot\rangle$
denotes the thermal average. It is found that
\begin{eqnarray}
\Theta & = & \exp\left[-\frac{1}{2}\sum_{k}\left(\frac{g_{k}}{\omega_{k}}\right)^{2}\coth\left(\frac{1}{2}\beta\omega_{k}\right)\right].\label{eq:theta}
\end{eqnarray}
Given a bath spectral density $J(\omega)=\sum_{k}g_{k}^{2}\delta(\omega-\omega_{k})$,
$\Theta$ can be rewritten as
\begin{equation}
\Theta=\exp\left[-\frac{1}{2}\int d\omega\frac{J(\omega)}{\omega^{2}}\coth\left(\frac{1}{2}\beta\omega\right)\right].
\end{equation}

Through out this paper, we consider the total initial state in the
original basis as
\begin{equation}
\rho_{\mathrm{tot}}(0)=\rho_{\mathrm{c}}\otimes|0\rangle\langle0|_{\mathrm{q}}
\otimes\rho_{\mathrm{b}},\label{eq:initial}
\end{equation}
where $\rho_{\mathrm{c}}$ is the reduced density matrix of the
chromophore, $|0\rangle\langle0|_{\mathrm{q}}$ denotes the spin ``up''
state of the qubit, and $\rho_{\mathrm{b}}=\exp(-\beta H_{\mathrm{b}})/Z$
is the thermalized phonon state in the original representation before polaron transformation.
Here $Z=\mathrm{Tr}\left[\exp(-\beta H_{\mathrm{b}})\right]$ is the partition
function and $\beta=1/(k_{\mathrm{b}}T)$, with $T$ the temperature
and $k_{\mathrm{b}}$ the Boltzmann constant. In the paper we
set $k_{\mathrm{b}}=1$.

In the interaction picture, the time dependent interaction Hamiltonian
can be expressed by $\tilde{H}_{\mathrm{I}}(t)=e^{i\left(\tilde{H}_{0}+H_{\mathrm{b}}\right)t}
\tilde{H}_{\mathrm{I}}e^{-i\left(\tilde{H}_{0}+H_{\mathrm{b}}\right)t}$.
After some algebra, we have
\begin{equation}
\tilde{H}_{\mathrm{I}}(t)=-\frac{\Delta}{2}\left[\sigma_{+1}(t)D(t)+\sigma_{-1}(t)D^{\dagger}(t)\right],
\end{equation}
in which the time dependent operator $\sigma_{\pm1}(t)$ reads $\sigma_{\pm1}(t)=e^{i\tilde{H}_{0}t}\sigma_{\pm1}e^{-i\tilde{H}_{0}t}$
with $\sigma_{\pm1}=(\sigma_{x1}\pm i\sigma_{y1})/2$ being the effective
raising (lowering) operator of the qubit. In the mean time, the reservoir
correlated operator $D(t)$ has a form of $D(t)=e^{B(t)}-\Theta$, where $B(t)=\sum_{k}(b_{k}^{\dagger}e^{i\omega_{k}t}-b_{k}e^{-i\omega_{k}t}){g_{k}}/{\omega_{k}}$.

The time evolution of the quasi-particle described by the effective Hamiltonian~(\ref{eq:H_0}) can
be solved using the time convolutionless polaron master equation~\cite{Breuer09,Wu13,Kolli11}.
In the polaron representation, assuming that $\tilde{\rho}_{\mathrm{cq}}$
is the reduced density matrix of the quasi-particle, then the
quantum master equation can be expressed by~\cite{Kolli11,Jang1,Jang2,Wu13}
\begin{eqnarray}
 &  & \partial_{t}\tilde{\rho}_{\mathrm{cq}}(t)+i\left[\tilde{H}_{0},\tilde{\rho}_{\mathrm{cq}}(t)\right]\nonumber \\
 & = & -i e^{-i\tilde{H}_{0}t}\mathrm{Tr}_{\mathrm{b}}\left\{ \left[\tilde{H}_{\mathrm{I}}(t),\mathcal{Q}\tilde{\rho}_{\mathrm{tot}}(0)\right]\right\} e^{i\tilde{H}_{0}t}\nonumber \\
 &  & -\int_{0}^{t}dse^{-i\tilde{H}_{0}t}\mathrm{Tr}_{\mathrm{b}}\left\{ \left[\tilde{H}_{\mathrm{I}}(t),\left[\tilde{H}_{\mathrm{I}}(s),\mathcal{Q}\tilde{\rho}_{\mathrm{tot}}(0)\right]\right]\right\} e^{i\tilde{H}_{0}t}\nonumber \\
 &  & -\int_{0}^{t}ds\mathrm{Tr}_{\mathrm{b}}\left\{ \left[\tilde{H}_{\mathrm{I}}(0),\left[\tilde{H}_{\mathrm{I}}(s-t),\tilde{\rho}_{\mathrm{cq}}(t)
 \otimes\rho_{\mathrm{b}}\right]\right]\right\} ,\label{eq:ME}
\end{eqnarray}
where
\begin{equation}
\mathcal{Q}\tilde{\rho}_{\mathrm{tot}}(0)=\tilde{\rho}_{\mathrm{tot}}(0)-\mathrm{Tr}_{\mathrm{b}}
\left[\tilde{\rho}_{\mathrm{tot}}(0)\right]\otimes\rho_{\mathrm{b}},
\end{equation}
with $\tilde{\rho}_{\mathrm{tot}}(0)=e^{U}\rho_{\mathrm{tot}}(0)e^{-U}$ being
the density matrix of the total ensemble in the polaron representation.
The general definition of the operator $\mathcal{Q}$ is given by~\cite{breuer_book}
\begin{equation}
\mathcal{Q}\rho:=\rho-\mathrm{Tr}_{\mathrm{b}}(\rho)\otimes\rho_{\mathrm{b}}.\label{eq:q}
\end{equation}
Taking into account the initial state (\ref{eq:initial}), one has
\begin{equation}
\mathcal{Q}\tilde{\rho}_{\mathrm{tot}}(0)=\rho_{\mathrm{c}}\otimes|0\rangle\langle0|
_{\mathrm{q}}\otimes\left(e^{B/2}\rho_{\mathrm{b}}e^{-B/2}-\rho_{\mathrm{b}}\right).
\end{equation}

In this model, with respect to the chromophore dynamics, it is found
that $\rho_{\mathrm{c,fix}}=|0\rangle\langle0|_{\mathrm{c}}$
is a dynamical fixed point, i.e., it does not evolve
with the passage of time, for which we will give a short proof here. Taking
$|0\rangle\langle0|_{\mathrm{c}}$ as the initial state of the chromophore,
in the Schr\"{o}dinger picture, we have
\begin{equation}
\rho_{\mathrm{c}}(t)=\mathrm{Tr}_{\mathrm{qb}}\left(e^{-i\tilde{H}t}|0\rangle\langle0|
_{\mathrm{c}}\otimes|0\rangle\langle0|_{\mathrm{q}}\otimes\rho_{\mathrm{b}}e^{i\tilde{H}t}
\right),\label{eq:rho_c}
\end{equation}
where the trace is taken over the subspaces of the qubit and the environment.
Based on the expression of $\tilde{H}$ and the fact that $|0\rangle$
is the eigenstate of $\sigma_{z}=|0\rangle\langle0|-|1\rangle\langle1|$,
one can find that
\begin{equation*}
\tilde{H}|0\rangle\langle0|_{\mathrm{c}}\otimes|0\rangle\langle0|_{\mathrm{q}}\otimes\rho_{\mathrm{b}}\tilde{H}
=|0\rangle\langle0|_{\mathrm{c}}\otimes\tilde{H}_{\mathrm{qb}}\left(|0\rangle\langle0|_{\mathrm{q}}
\otimes\rho_{\mathrm{b}}\right)\tilde{H}_{\mathrm{qb}},
\end{equation*}
where $\tilde{H}_{\mathrm{qb}}=\tilde{H}_{0\mathrm{qb}}+\tilde{H}_{\mathrm{I}}+H_{\mathrm{b}}$,
and $\tilde{H}_{0\mathrm{qb}}=\frac{\omega_{0}}{2}+\frac{\epsilon}{2}\sigma_{z1}-\frac{\Delta}{2}\sigma_{x1}\Theta
+\frac{a}{2}\sigma_{z1}-\sum_{k}\frac{g_{k}^{2}}{4\omega_{k}}.$
It should be noted that though the notations in this Hamiltonian seem similar to
those in $\tilde{H}$, the operators in $\tilde{H}_{\mathrm{qb}}$
are actually confined to the subspace of the qubit and the environment.
Then Eq.~(\ref{eq:rho_c}) can be rewritten as $\rho_{\mathrm{c}}(t)=|0\rangle\langle0|_{\mathrm{c}}\mathrm{Tr}_{\mathrm{qb}}(e^{-i\tilde{H}_{\mathrm{qb}}t}
|0\rangle\langle0|_{\mathrm{q}}\otimes\rho_{\mathrm{b}}e^{i\tilde{H}_{\mathrm{qb}}t})$.
Based on the cyclic permutation invariance of trace, it can be simplified as
\begin{equation}
\rho_{\mathrm{c}}(t)=|0\rangle\langle0|_{\mathrm{c}}.
\end{equation}
Thus, $|0\rangle\langle0|_{\mathrm{c}}$ is a dynamical fixed point
of the chromophore. More generally, as the interaction Hamiltonian (\ref{eq:inter_Hamil}) commutes with the density matrix of the chromophore, and the interaction between the chromophore and qubit is of $\sigma_{z}$ type, the dissipative process of the chromophore is through dephasing. Thus, all the diagonalized states of the chromophore are
fixed points. Dynamical fixed points, which are especially useful in the study of the decoherence-free subspace, will be utilized in this work to construct a new measure of non-Markovianity.

\section{Non-Markovianity} \label{sec:NM}

To quantify the non-Markovian behavior, several measures have been
proposed~\cite{Breuer09,Rivas10,Liu13,Lu10,Luo12,Chruscinski10}, among which the BLP measure~\cite{Breuer09} relates the non-Markovian
behavior to the backflow information from the reservoir to the system.
The BLP definition is based on the trace distance
\begin{equation}
D_{\mathrm{tr}}\left(\rho_{1},\rho_{2}\right):=\frac{1}{2}\mathrm{Tr}|\rho_{1}-\rho_{2}|,
\end{equation}
where $|O|=\sqrt{O^{\dagger}O}$. For a single qubit, it is known
that its density matrix can be expressed in the Bloch representation as
\begin{equation}
\rho=\frac{1}{2}\left(\openone+\bm{r}\cdot\bm{\sigma}\right),
\end{equation}
where $\bm{r}$ is the Bloch vector and $\bm{\sigma}=(\sigma_{x},\sigma_{y},\sigma_{z})^{\mathrm{T}}$
with $\sigma_{x,y,z}$ the Pauli matrix. In this representation, the
trace distance can be reduced to the Euclidean distance between the Bloch
vectors~\cite{Nielsen}
\begin{equation}
D_{\mathrm{tr}}(\rho_{1},\rho_{2})=\frac{1}{2}||\bm{r}_{1}-\bm{r}_{2}||.
\end{equation}
Here $||\cdot||$ is the Euclidean distance. $\bm{r}_{1}$ and $\bm{r}_{2}$
are the corresponding Bloch vectors of $\rho_{1}$ and $\rho_{2}$,
respectively.

The trace distance is monotonous when the system goes through the quantum channels, which can
be described by the completely positive and trace-preserving
maps. Physically, this monotonicity is explained intuitively by that
the information of distinguishability always flows from the system
to the environment when the system goes through a quantum channel.
Thus, the backflow of the information can be treated as a non-Markovian
behavior, or a memory effect in which the environment absorbs information
from the system and return some back to it, improving the distinguishability
of the system. The BLP non-Markovianity is defined based on such a mechanism:
\begin{equation}
\mathcal{N}_{\mathrm{BLP}}:=\underset{\rho_{1,2}(0)}{\mathrm{max}}\int_{\sigma>0}dt\sigma
\left(t,\rho_{1,2}(0)\right),\label{eq:BLP}
\end{equation}
where
\begin{equation}
\sigma\left(t,\rho_{1,2}(0)\right):=\frac{d}{dt}D_{\mathrm{tr}}\left(\rho_{1}(t),\rho_{2}(t)\right)
\end{equation}
is the time derivative of the trace distance during the evolution.
The maximum is taken over all pairs of initial states $\rho_{1}(0)$
and $\rho_{2}(0)$. According to the definition, it can be found that $\mathcal{N}_{\mathrm{BLP}}\geq0$.
For a Markovian process, the non-Markovianity is zero, i.e., $\mathcal{N}_{\mathrm{BLP}}=0$.

Dynamical fixed points may adopt various forms in physical systems, including states that are thermalized and ones within a decoherence-free subspace of a quantum system~\cite{DFS,DFS_1,DFS_2,DFS_3}. An alternative definition of non-Markovianity is introduced below based on the trace distance in the presence of dynamical fixed points
\begin{equation}
\mathcal{N}_{\mathrm{fix}}=\max_{\rho(0)}\int_{\sigma_{\mathrm{fix}>0}}dt\sigma_{\mathrm{fix}}
\left(t,\rho(0)\right),\label{eq:fix_NM}
\end{equation}
where
\begin{equation}
\sigma_{\mathrm{fix}}\left(t,\rho(0)\right)
:=\frac{d}{dt}D_{\mathrm{tr}}\left(\rho(t),\rho_{\mathrm{fix}}\right).
\end{equation}
Here $\rho_{\mathrm{fix}}$ is the matrix for any dynamical fixed point,
which satisfies the equation $\partial_{t}\rho_{\mathrm{fix}}=0$,
and the maximum is taken over all the initial state $\rho(0)$. It
is easy to verify that $\mathcal{N}_{\mathrm{fix}}\geq0$, and $\mathcal{N}_{\mathrm{fix}}=0$
for Markovian dynamics. This measure can be treated as a special form
of the BLP measure as they both rely on the same mechanism, i.e.,
monotonicity of the trace distance under the completely positive and trace-preserving
maps. The value of $\mathcal{N}_{\mathrm{fix}}$ may be equal or less
than that of $\mathcal{N}_{\mathrm{BLP}}$. However, most of the physical
information is contained in the variation of the non-Markovianity,
not its absolute value. Therefore, despite that $\mathcal{N}_{\mathrm{fix}}$
does not contain as many pairs of initial states as $\mathcal{N}_{\mathrm{BLP}}$,
it is still capable to describe the system behavior.

%------------------Figure 2------------------
\begin{figure}[tp]
\includegraphics[width=7.5cm]{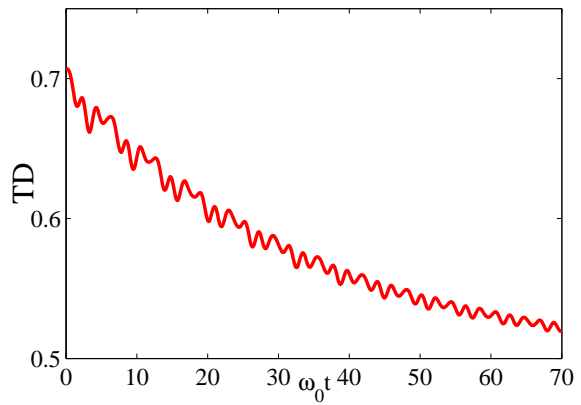}
\caption{\label{fig:TD} The dynamics of the trace distance of
the chromophore. The initial states are $\bm{r}_{\mathrm{c1}}=\left(1,0,0\right)^{\mathrm{T}}$
and $\bm{r}_{\mathrm{c,fix}}=\left(0,0,1\right)^{\mathrm{T}}$. The
parameters are set as $\epsilon=\omega_{0}$, $\Delta=0.8\omega_{0}$, $a=2\omega_{0}$, and $T=0.5\omega_{0}$.}
\end{figure}
%-----------------------------------------------

Moreover, compared with BLP measure, the $\mathcal{N}_{\mathrm{fix}}$
measure has a numerical advantage. In principle, there are an infinite
number of states in the Hilbert space. To carry out the numerical
calculation of Eq.~(\ref{eq:BLP}), one has to sample a finite number
of them. Assuming that this number is $N$, to take over all pairs
of initial states, it generally requires $N(N-1)/2$ times of calculations.
Utilizing Eq.~(\ref{eq:fix_NM}), the calculation number is only
$N$. Therefore, for some complex dynamics with fixed points in it,
Eq.~(\ref{eq:fix_NM}) provides an efficient algorithm with a $\mathcal{O}(N)$ numerical advantage.

Since the maximization in $\mathcal{N}_{\mathrm{fix}}$ is not taken over all pairs of initial states in the Hilbert space, this measure does not capture the full information on non-Markovianity. This incompleteness of information on non-Markovianity is a trade-off of numerical advantage. However, since the non-Markovianity is a property of system dynamics, it should be independent of the initial states in principle. As a matter of fact, a dynamics can be called
non-Markovian dynamics if the trace distance of any two states has the increasing behavior during the evolution. Thus, the size of the set in which the maximization is performed is not a decisive factor on the issue of non-Markovianity. Admittedly, Eq.~(21) does not give a maximizing set as large as that of the BLP measure. We take a dynamical fixed point as the datum line because it does not evolve over the considered timescale. We then calculate the trace distance between this fixed point and all the states in Hilbert space, indicating that the dynamical information of all initial states are involved in this measure. This explains why Eq.~(21) is capable to quantify the behaviors of non-Markovianity.
Nevertheless, in some extreme mathematical cases where the numerical advantage of Eq.~(21) is not obvious, the BLP measure would be a better choice.

As we discussed in Sec.~\ref{sec:MandD}, the state $\rho_{\mathrm{c,fix}}=|0\rangle\langle0|_{\mathrm{c}}$ is
a dynamical fixed point of the chromophore. Thus, it is convenient to use Eq.~(\ref{eq:fix_NM}) in
our model to describe the non-Markovian behavior of the chromophore.

\section{Discussion} \label{sec:Discussion}

In this section, we will apply the non-Markovianity measure~(\ref{eq:fix_NM})
in the model given in Sec.~\ref{sec:MandD}. With this measure, we
further discuss the non-Markovian behavior of the chromophore, as
well as  the chromophore-qubit pair, in the presence of the phonon bath.
The effects of the temperature and the coupling strength
between the chromophore and qubit on the non-Markovianity will be discussed.

Generally, the expectation value of an observable $A$ of the chromophore-qubit pair can be written as $\langle A\rangle=
\mathrm{Tr}_{\mathrm{cqb}}\left[A\rho_{\mathrm{tot}}(t)\right]$,
where $\rho_{\mathrm{tot}}(t)$ is the total density matrix in the
original representation including the chromophore, the qubit and the
phonon bath. The subscript c, q and b represent the subspaces of the chromophore,
the qubit and the bath, respectively. Using the inverse polaron transformation
and inserting $\mathcal{P}+\mathcal{Q}$ into the expression, one
can obtain the expectation of $A$ as~\cite{Wu13,Kolli11}
\begin{equation}
\langle A\rangle=\langle A\rangle_{\mathrm{rel}}+\langle A\rangle_{\mathrm{irrel}},
\end{equation}
where $\langle A\rangle_{\mathrm{rel}}$ is the relevant part, which
can be written as
\begin{equation}
\langle A\rangle_{\mathrm{rel}}=\mathrm{Tr}_{\mathrm{cq}}\left[\tilde{\rho}_{\mathrm{cq}}(t)\mathrm{Tr}_{\mathrm{b}}
\left(e^{U}Ae^{-U}\rho_{\mathrm{b}}\right)\right],\label{eq:rel}
\end{equation}
and the irrelevant part $\langle A\rangle_{\mathrm{irrel}}$ reads
\begin{equation}
\langle A\rangle_{\mathrm{irrel}}=\mathrm{Tr}_{\mathrm{cqb}}
\left[e^{U}Ae^{-U}\mathcal{Q}\tilde{\rho}_{\mathrm{tot}}(t)\right].\label{eq:irrel}
\end{equation}

%-----------------------Figure 3----------------------
\begin{figure}
\includegraphics[width=7cm]{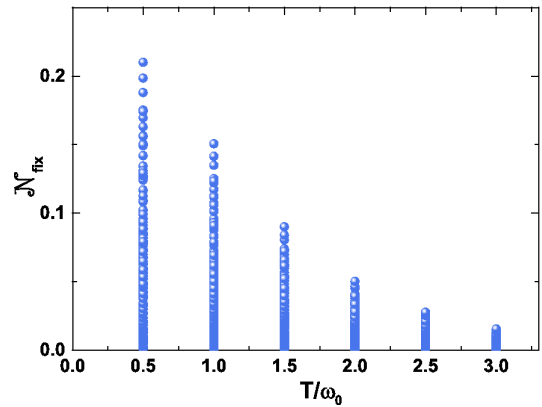}
\caption{\label{fig:temperature} The variation of non-Markovianity
$\mathcal{N}_{\mathrm{fix}}$ as a function of the normalized temperature $T/\omega_{0}$.
The parameters are set as $\epsilon=\omega_{0}$, $\Delta=0.8\omega_{0}$, and
$a=4\omega_{0}$. The spectral density and the relevant parameters
are the same as those in Fig.~\ref{fig:TD}.}
\end{figure}
%--------------------------------------------------------

When $\left[A,U\right]=0$, the irrelevant part can be simplified
into $\langle A\rangle_{\mathrm{irrel}}=\mathrm{Tr}_{\mathrm{cq}}\left\{ A\,\mathrm{Tr}_{\mathrm{b}}\left[\mathcal{Q}\tilde{\rho}_{\mathrm{tot}}(t)\right]\right\} $.
From the definition of $\mathcal{Q}$ in Eq.~(\ref{eq:q}), it is
easy to see that $\mathrm{Tr}_{\mathrm{b}}[\mathcal{Q}\tilde{\rho}_{\mathrm{tot}}(t)]=0$.
Thus, for those observables that commute with the generator of the
polaron transformation, their expectations contains only the relevant part, i.e.,
\begin{equation}
\langle A\rangle=\mathrm{Tr}_{\mathrm{cq}}\left[A\tilde{\rho}_{\mathrm{cq}}(t)\right].\label{eq:expectation}
\end{equation}

\subsection{The chromophore}

With above preliminary knowledge, we will try to reproduce the dynamical
information of the chromophore in the original basis. For the chromophore-qubit pair, its density matrix in the original
basis can always be decomposed into the form~\cite{Fano83,Luo08}
\begin{equation}
\rho_{\mathrm{cq}}=\frac{1}{4}\left(\openone+\bm{r}_{\mathrm{c}}\cdot\bm{\sigma_{0}}
+\bm{r}_{\mathrm{q}}\cdot\bm{\sigma_{1}}+\bm{m}\cdot\bm{\sigma_{m}}\right),\label{eq:2qubit_rho}
\end{equation}
where $\bm{r}_{\mathrm{c}}$ and $\bm{r}_{\mathrm{q}}$ are the Bloch
vectors of the chromophore and qubit, respectively, while $\bm{\sigma}_{i}=(\sigma_{xi},\sigma_{yi},\sigma_{zi})^{\mathrm{T}}$
for $i=0,1.$ and $\bm{\sigma_{m}}=\left(\sigma_{x}\otimes\sigma_{x},\sigma_{y}\otimes\sigma_{y},
\sigma_{z}\otimes\sigma_{z}\right)^{\mathrm{T}}$.
Through some straightforward calculation, one can find
\begin{equation}
\bm{r}_{\mathrm{c}}=\left(\langle\sigma_{x0}\rangle,\langle\sigma_{y0}\rangle,\langle\sigma_{z0}
\rangle\right)^{\mathrm{T}},
\end{equation}
where the expectation $\langle\sigma_{i0}\rangle=\mathrm{Tr}_{\mathrm{cq}}(\sigma_{i0}\rho_{\mathrm{cq}})$
for $i=x,y,z$. As $[\sigma_{i0},U]=0$, based on Eq.~(\ref{eq:expectation}),
the expectation of $\sigma_{i0}$ in original representation is the
same as that in the polaron representation, namely,
\begin{equation}
\langle\sigma_{i0}\rangle=\mathrm{Tr}_{\mathrm{cq}}\left[\sigma_{i0}\tilde{\rho}_{\mathrm{cq}}(t)\right].
\end{equation}
In this way, we can reproduce the dynamical information of the chromophore.
Using the expression of Bloch vector $\bm{r}_{\mathrm{c}}$, the trace distance between $\bm{r}_{\mathrm{c}}$
and the fixed point $\bm{r}_{\mathrm{c,fix}}$ can be written as
\begin{equation}
D_{\mathrm{tr}}=\frac{1}{2}\sqrt{\left(1-\langle\sigma_{z0}\rangle\right)^{2}
+\langle\sigma_{x0}\rangle^{2}+\langle\sigma_{y0}\rangle^{2}}.
\end{equation}
The equivalent expression using the elements of the density matrix is $D_{\mathrm{tr}}=\sqrt{\rho_{\mathrm{c},22}^{2}(t)+|\rho_{\mathrm{c},12}(t)|^{2}}$.

%---------------------Figure 4-----------------------
\begin{figure}
\includegraphics[width=7.8cm]{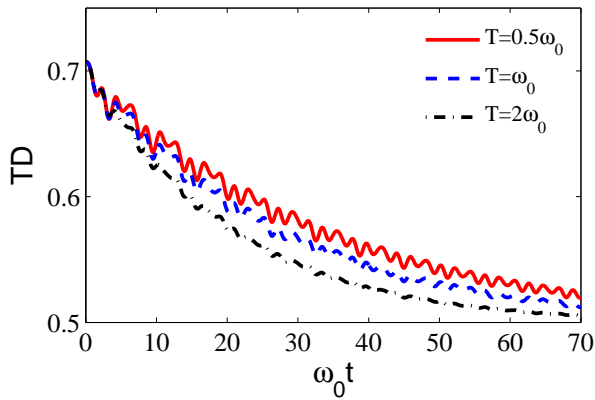}\caption{\label{fig:temperature_dis} The time evolution of trace
distance under different temperatures. The values of temperature $T$
are $0.5\omega_{0}$, $\omega_{0}$ and $2\omega_{0}$ for the red
solid line, the blue dashed line and black dashed dotted line, respectively.
The parameters are set as $\epsilon=\omega_{0}$, $\Delta=0.8\omega_{0}$,
and $a=2\omega_{0}$. The Bloch vectors of the initial states are $\bm{r}_{\mathrm{c}1}=(1,0,0)^{\mathrm{T}}$
and $\bm{r}_{\mathrm{c,fix}}=(0,0,1)^{\mathrm{T}}$.}
\end{figure}
%--------------------------------------------------------

Figure~\ref{fig:TD} displays the time evolution of the trace distance
of the chromophore.
The initial states are chosen as $r_{\mathrm{c1}}=\left(1,0,0\right)^{\mathrm{T}}$
and $\bm{r}_{\mathrm{c,fix}}=\left(0,0,1\right)^{\mathrm{T}}$. Here
$\bm{r}_{\mathrm{c,fix}}$ is the Bloch vector of the dynamical fixed
point of the chromophore: $\rho_{\mathrm{c,fix}}=|0\rangle\langle0|_{\mathrm{c}}$.
The parameters are set as $\epsilon=\omega_{0}$, $\Delta=0.8\omega_{0}$,
$a=2\omega_{0}$, and $T=0.5\omega_{0}$. A super-Ohmic spectral density $J(\omega)=\kappa\omega_{\mathrm{ph}}^{-2}\omega^{3}\exp(-\omega/\omega_{c})$
is taken with the characteristic phonon frequency $\omega_{\mathrm{ph}}=\omega_{0}$,
the cutoff frequency $\omega_{c}=4\omega_{0}$, the coupling strength
$\kappa=0.1\omega_{0}$.
It is found in Fig.~\ref{fig:TD} that the trace distance has an
oscillating component during its evolution. The descending trend of
the trace distance is due to the dissipative effect of the environment,
indicating an information flow from the system to the environment.
However, this information flow is by no means unidirectional. The
oscillation of the trace distance demonstrates the back flow
of information from the environment to the system, pointing to the
non-Markovian behavior in the dynamics of the chromophore.

As part of the system-environment correlation, the non-Markovianity
must be affected by the temperature of the environment, as discussed in the literatures~\cite{Clos,Vasile14,breuer_book}.
Here we also examine the influence of temperature on the non-Markovianity.
Figure~\ref{fig:temperature} shows the behavior of non-Markovianity
$\mathcal{N}_{\mathrm{fix}}$ as a function of the temperature. The
parameters in this figure are set as $ $$\epsilon=\omega_{0}$, $\Delta=0.8\omega_{0}$,
and $a=4\omega_{0}$. The spectral density and the relevant parameters
are the same as those in Fig.~\ref{fig:TD}. We have taken over more
than 1000 initial states for each temperature point in this plot.
Figure~\ref{fig:temperature} demonstrates that the increase of the temperature
can suppress the non-Markovianity.

%---------------------------Figure 5-------------------------------
\begin{figure}
\includegraphics[width=7cm]{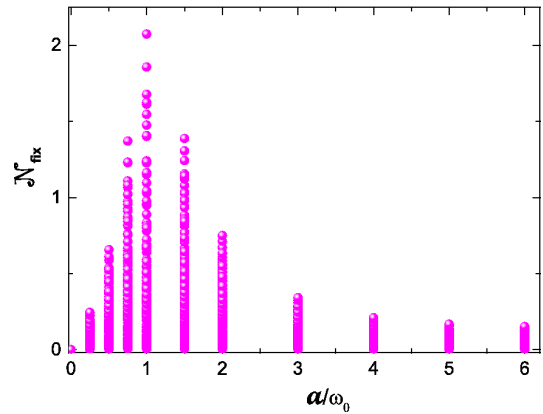}\caption{\label{fig:couplig} The variation of non-Markovianity
$\mathcal{N}_{\mathrm{fix}}$ as a function of the normalized coupling
strength $a/\omega_{0}$. The parameters in this figure are set as
$\epsilon=\omega_{0}$, $\Delta=0.8\omega_{0}$, and $T=0.5\omega_{0}$.
The spectral density and the relevant parameters are the same as those
in Fig.~\ref{fig:TD}.}
\end{figure}
%---------------------------------------------------------------------

To further probe the temperature effect, we plot the evolution of
the trace distance with specific initial states $\bm{r}_{\mathrm{c1}}=(1,0,0)^{\mathrm{T}}$
and $\bm{r}_{\mathrm{c,fix}}=(0,0,1)^{\mathrm{T}}$, as shown in Fig.~\ref{fig:temperature_dis}.
It is found that the increasing temperature can speed up the decay
of trace distance, as well as suppress its oscillation. More phonons
are excited with the rise of temperature, leading to the speed up of decoherence
of chromophore via the probe qubit. The acceleration of the decoherence
will reduce the oscillation amplitudes of the coherence
of the density matrix of chromophore and increase its decay rate. Then, based on the equation
$D_{\mathrm{tr}}^{2}=\rho_{\mathrm{c},22}^{2}(t)+|\rho_{\mathrm{c,12}}(t)|^{2}$,
one can see that the speed up of decoherence results in a faster decay
of the trace distance, and its oscillating amplitude, causing a further
reduction of non-Markovianity and therefore a suppression of the
information backflow. The fact that the increase of temperature can suppress the
non-Markovianity has also appeared in other systems~\cite{Clos,Vasile14,breuer_book}.
This coincidence indicates that Eq.~(\ref{eq:fix_NM}) is qualified to capture the non-Markovian
behavior of dynamics in this system.

Another important parameter affecting the system-environment
correlation is the coupling strength $a$ between the chromophore
and qubit, which is proportional to the angular orientation parameter $\eta$
of the dipole-dipole interaction and inversely proportional to the cube
of the distance $r$ between the chromophore and qubit~\cite{Brown},
i.e., $a\propto\eta/r^{3}$. In previous studies where the dynamical processes
are solved by the perturbation methods, it is difficult to gauge
the effect of the coupling strength on non-Markovianity due to the weak
coupling constraint. In our cases, this constraint is reflected by the parameter $\kappa$ in the spectral density, i.e., $\kappa$ cannot be arbitrarily large. However, the coupling between the chromophore and qubit can be chosen in a wide range, without affecting the accuracy of the time-convolutionless master equation, allowing a study on the non-Markovianity in the strong coupling regime.

%------------------------------Figure 6-----------------------------
\begin{figure}
\includegraphics[width=8.5cm]{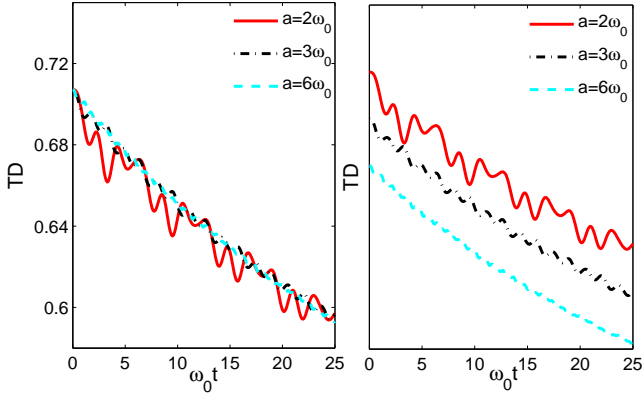}\caption{\label{fig:TD_a} The time evolution of the trace distance
with three values of the coupling strength $a$. Left Panel: the red solid line, the black
dashed dotted line and the blue dashed line represent the trace distances
with $a=2\omega_{0}$, $a=3\omega_{0}$ and $a=6\omega_{0}$, respectively.
Other parameters are the same as Fig.~\ref{fig:couplig}. Right Panel: the solid red (dashed blue) lines in
Left Panel is shifted upward
(downward) to better distinguish the three cases. }
\end{figure}
%-----------------------------------------------------------------------

Figure~\ref{fig:couplig} shows the variation of the non-Markovianity
$\mathcal{N}_{\mathrm{fix}}$ as a function of coupling parameter
$a$. Other parameters in this figure are set as $\epsilon=\omega_{0}$,
$\Delta=0.8\omega_{0}$, and $T=0.5\omega_{0}$. The spectral density
and the relevant parameters are the same as those in the previous
figures. Similarly with Fig.~\ref{fig:temperature}, we also take over
more than 1000 initial states for each value of $a$. When $a=0$,
the non-Markovianity vanishes. This is because when the chromophore
is isolated, its evolution turns out to be unitary and the trace distance
is unchanged for a unitary transformation. In other words, there is
no information exchange between the chromophore and qubit in this
case. In the weak coupling regime, the increase of $a$ can enhance
the non-Markovianity. This enhancement effect is well known in the
community and is one of the main reasons why the Markovian approximation
is applicative in the weak coupling regime. However, when $a$ is
large, the effect is totally contrary. In this regime, the increase
of $a$ will suppress the non-Markovianity. In our case, the maximum non-Markovianity
is obtained around $a=\omega_{0}$, a value that may change if system parameters vary.

%------------------------------Figure 7-----------------------------
\begin{figure}
\includegraphics[width=7.5cm]{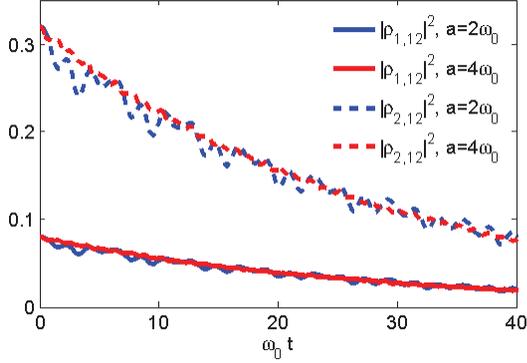}\caption{\label{fig:evolution}(Color online) Time evolution of the square norm of the off-diagonal element of chromophore's reduced density matrix. Two initial states, labeled by 1 and 2, are considered: $\rho_{1,11}=0.5$, $\rho_{1,12}=0.2$ and $\rho_{2,11}=0.7$, $\rho_{2,12}=0.4$. The solid (dashed) blue and red lines represent $|\rho_{1,12}|^2$ ($|\rho_{2,12}|^2$) for $a=2\omega_{0}$ and $a=4\omega_{0}$, respectively. Other parameters are the same as in Fig.~\ref{fig:couplig}.}
\end{figure}
%-----------------------------------------------------------------------

To obtain a deeper understanding of how the system-bath coupling
suppresses the non-Markovianity in the strong coupling regime,
we plot the time evolution of the trace
distances for three values of $a$ in Fig.~\ref{fig:TD_a}. The
red solid line, the black dashed dotted line and the blue dashed line
represent the trace distances with $a=2\omega_{0}$, $a=3\omega_{0}$
and $a=6\omega_{0}$, respectively. Other parameters are the same
as those in Fig.~\ref{fig:couplig}. The black and blue lines are shifted
downward in the right panel to better distinguish the curves. It is found that
the increase of coupling strength will not affect the decay rate of
the trace distance, but does affect its oscillation amplitudes. This is because
in this coupling regime, the behavior of the system trends to a way similar to
the overdamped behavior. The oscillation of the off-diagonal element of the density matrix
is suppressed.

To show this, we plot as a function of time the square norm of the off-diagonal element of chromophore's reduced density matrix in Fig.~\ref{fig:evolution}. Two
specific initial states, labeled by 1 and 2, are considered: $\rho_{1,11}=0.5$, $\rho_{1,12}=0.2$ and $\rho_{2,11}=0.7$, $\rho_{2,12}=0.4$. The solid (dashed) blue and red lines represent $|\rho_{1,12}|^2$ ($|\rho_{2,12}|^2$) for $a=2\omega_{0}$ and $a=4\omega_{0}$, respectively. Other parameters are the same as in Fig.~\ref{fig:couplig}. It is shown in Fig.~\ref{fig:evolution} that the oscillation amplitude of the off-diagonal elements decreases with increasing $a$, but the decay rate stays constant.
As this oscillation attenuation happens to all states but the
fixed points in Hilbert space, any measure based
on the trace distance will exhibit this behavior, including
the BLP measure. It is thus an intrinsic property
of the system non-Markovianity.

\subsection{The chromophore-qubit pair}

In this subsection, we look into the non-Markovian behavior
of the chromophore-qubit pair in the presence of the bath.
First, its dynamics can be obtained by calculating the vectors
$\bm{r}_{\mathrm{c}}$, $\bm{r}_{\mathrm{q}}$ and $\bm{m}$ in Eq.~(\ref{eq:2qubit_rho}).
The expression of $\bm{r}_{\mathrm{c}}$ has been given in the previous
subsection. Now we will focus on $\bm{r}_{\mathrm{q}}$ and $\bm{m}$.
Based on Eq.~(\ref{eq:2qubit_rho}), one can easily find that
\begin{align}
\bm{r}_{\mathrm{q}} & =\left(\langle\sigma_{x1}\rangle,\langle\sigma_{y1}\rangle,\langle\sigma_{z1}\rangle\right)^{\mathrm{T}},\\
\bm{m} & =\left(\langle\sigma_{x}\otimes\sigma_{x}\rangle,\langle\sigma_{y}\otimes\sigma_{y}\rangle,\langle\sigma_{z}\otimes\sigma_{z}\rangle\right)^{\mathrm{T}}.
\end{align}

As some of the operators above do not commute with the generator
$U$, the corresponding irrelevant parts of the expectations do not
vanish. In general, this irrelevant contribution is hard to obtain
because of the complex form of $\mathcal{Q}\tilde{\rho}_{\mathrm{tot}}(t)$.
For simplicity, one can choose the zeroth order of the $\mathcal{Q}\tilde{\rho}_{\mathrm{tot}}(t)$
as an approximation~\cite{Kolli11,Wu13}. In Ref.~\cite{Kolli11},
the zeroth order approximation was taken in the Schr\"{o}dinger picture,
i.e.,
\begin{equation*}
\langle A\rangle_{\mathrm{irrel}}^{\mathrm{S}}=\mathrm{Tr}_{\mathrm{cq}}\!\left[A\rho_{\mathrm{cq}}\left(0\right)\right]-\mathrm{Tr}_{\mathrm{cq}}\!\left[\tilde{\rho}_{\mathrm{cq}}(0)\mathrm{Tr}_{\mathrm{b}}\left(e^{U}Ae^{-U}\rho_{\mathrm{b}}\right)\right],
\end{equation*}
while the approximation can also be made in the interaction picture,
\begin{equation*}
\langle A\rangle_{\mathrm{irrel}}^{\mathrm{I}}=\mathrm{Tr}_{\mathrm{cqb}}\!\left[e^{i(\tilde{H}_{0}+H_{\mathrm{b}})t}e^{U}Ae^{-U}e^{-i(\tilde{H}_{0}+H_{\mathrm{b}})t}\mathcal{Q}\tilde{\rho}_{\mathrm{tot,I}}(0)\right],
\end{equation*}
where $\tilde{\rho}_{\mathrm{tot,I}}$ is the density matrix of the
total ensemble in the interaction picture. When $[A,U]=0$, the irrelevant
part vanishes, then the two approximations are actually the same.

%--------------------------Figure 8--------------------------
\begin{figure}
\includegraphics[width=7.2cm]{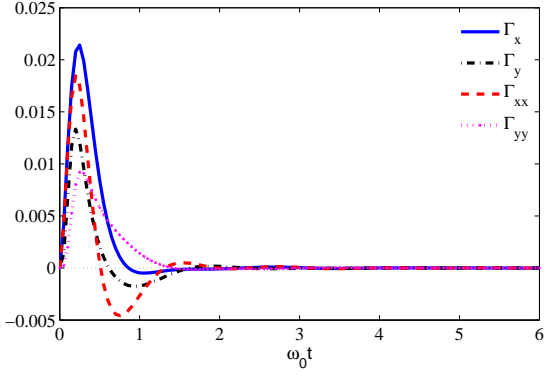}
\caption{\label{fig:different} The variations of $\Gamma_{i}$ and
$\Gamma_{ii}$ for $i=x,y$ as a function of time. Here $\Gamma_{i}=\langle\sigma_{i1}\rangle_{\mathrm{irrel}}^{\mathrm{S}}-\langle\sigma_{i1}\rangle_{\mathrm{irrel}}^{\mathrm{I}}$
and $\Gamma_{ii}=\langle\sigma_{i}\otimes\sigma_{i}\rangle_{\mathrm{irrel}}^{\mathrm{S}}-\langle\sigma_{i}\otimes\sigma_{i}\rangle_{\mathrm{irrel}}^{\mathrm{I}}$.
The parameters in this figure are set as $\epsilon=\omega_{0}$, $\Delta=0.8\omega_{0}$,
$T=0.5\omega_{0}$, and $a=4\omega_{0}$. The spectral density and the
relevant parameters are the same as those in Fig.~\ref{fig:TD}.
The blue solid, black dashed dotted, red dashed and pink dotted lines
represent $\Gamma_{x}$, $\Gamma_{y}$, $\Gamma_{xx}$ and $\Gamma_{yy}$,
respectively.}
\end{figure}
%----------------------------------------------------------------

Since the density matrix in the original basis is our main concern,
we calculate the differences of the expectations $\langle\sigma_{i1}\rangle$ and
$\langle\sigma_{i}\otimes\sigma_{i}\rangle$ for $i=x,y$ by performing
these two approximations. Through some straightforward calculation,
it can be found that the irrelevant part
$\langle\sigma_{x1}\rangle_{\mathrm{irrel}}^{\mathrm{I}}
=2\mathrm{Re}[\chi(t)\langle\sigma_{+1}(t)\rangle_{\rho_{\mathrm{cq}}(0)}]$,
where the time-dependent function $\chi(t)$ is defined as $\chi(t):=\Theta\{\exp[\sum_{k}i\sin(\omega_{k}t)g_{k}^{2}/\omega_{k}^{2}]-1\}$,
$\mathrm{Re}(\cdot)$ denotes the real part,
and $\langle\cdot\rangle_{\rho_{\mathrm{cq}(0)}}:=\mathrm{Tr}_{\mathrm{cq}}[\cdot\rho_{\mathrm{cq}}(0)]$.
Here $\rho_{\mathrm{cq}}(0)=\rho_{\mathrm{c}}\otimes |0\rangle\langle0|_{\mathrm{q}}$.
Similarly, one can obtain that $\langle\sigma_{y1}\rangle_{\mathrm{irrel}}^{\mathrm{I}}
=2\mathrm{Im}[\chi(t)\langle\sigma_{+1}(t)\rangle_{\rho_{\mathrm{cq}}(0)}]$.
As the polaron transformation generator $U$ has nothing to do with
the subspace of the chromophore, one arrives at $\langle\sigma_{x}\otimes\sigma_{x}\rangle_{\mathrm{irrel}}^{\mathrm{I}}
=2\mathrm{Re}[\chi(t)\langle\{\sigma_{x}\otimes\sigma_{+}\}(t)\rangle_{\rho_{\mathrm{cq}}(0)}]$,
where $\{\sigma_{x}\otimes\sigma_{+}\}(t):=e^{i\tilde{H}_{0}t}(\sigma_{x}\otimes\sigma_{+})e^{-i\tilde{H}_{0}t}$
is the time dependent operator of $\sigma_{x}\otimes\sigma_{+}$ in
the interaction picture. It is therefore found that $\langle\sigma_{y}\otimes\sigma_{y}\rangle_{\mathrm{irrel}}^{\mathrm{I}}
=2\mathrm{Im}[\chi(t)\langle\{\sigma_{y}\otimes\sigma_{+}\}(t)\rangle_{\rho_{\mathrm{cq}}(0)}]$.
Utilizing the approximation in the Schr\"{o}dinger picture, and considering
the initial state~(\ref{eq:initial}), the irrelevant parts of the
expectations of these four operators all vanish. We will compare below the expectation values from these two approximations.

Denote $\Gamma_{i}=\langle\sigma_{i1}\rangle_{\mathrm{irrel}}^{\mathrm{S}}
-\langle\sigma_{i1}\rangle_{\mathrm{irrel}}^{\mathrm{I}}$
and $\Gamma_{ii}=\langle\sigma_{i}\otimes\sigma_{i}\rangle_{\mathrm{irrel}}^{\mathrm{S}}
-\langle\sigma_{i}\otimes\sigma_{i}\rangle_{\mathrm{irrel}}^{\mathrm{I}}$,
where $i=x,y$, as differences in operator expectation values between the
two approximations. In Fig.~\ref{fig:different}, $\Gamma_{i}$ and
$\Gamma_{ii}$ are plotted as a function of time, where the
parameters are set as $\epsilon=\omega_{0}$, $\Delta=0.8\omega_{0}$,
$T=0.5\omega_{0}$, and $a=4\omega_{0}$. The spectral density and the
relevant parameters are the same as those in Fig.~\ref{fig:TD}.
The initial Bloch vector for the chromophore is taken as $\bm{r}_{\mathrm{c}1}=(1,0,0)^{\mathrm{T}}$.
The blue solid, black dashed dotted, red dashed and pink dotted lines
represent $\Gamma_{x}$, $\Gamma_{y}$, $\Gamma_{xx}$ and $\Gamma_{yy}$,
respectively. It is found that the differences in the
the expectation values are small and short-lived, disappearing beyond $\omega_{0}t=2$. For convenience,
the approximation performed in the Schr\"{o}dinger picture is chosen
because the irrelevant part of the expectation values we are interested
in always vanish in this case.

Thus, for the zeroth order approximation, taking into account the initial state Eq.~(\ref{eq:initial}), the expectation value of operator A has the form
\begin{equation}
\langle A\rangle=\mathrm{Tr}_{\mathrm{cq}}\left[\tilde{\rho}_{\mathrm{cq}}(t)\mathrm{Tr}_{\mathrm{b}}\left(e^{U}Ae^{-U}
\rho_{\mathrm{b}}\right)\right].
\end{equation}
Here $A=\sigma_{x1}$, $\sigma_{y1}$, $\sigma_{x}\otimes \sigma_{x}$, or  $\sigma_{y}\otimes \sigma_{y}$. Density matrix dynamics can be obtained
for the chromophore-qubit pair in the original basis.
Keeping in mind the fact that $e^{U}\sigma_{x1}e^{-U}=\sigma_{x1}\cosh B+i\sigma_{y1}\sinh B$, and $\langle\sinh B\rangle=0$,
and utilizing the zeroth approximation, the expectation value
of $\sigma_{x1}$ in the original basis can be written as $\langle\sigma_{x1}\rangle=\Theta\mathrm{Tr}_{\mathrm{eq}}\left[\sigma_{x1}\tilde{\rho}_{\mathrm{cq}}(t)\right]$.
Similarly, one can obtain that $\langle\sigma_{y1}\rangle=\Theta\mathrm{Tr}_{\mathrm{eq}}\left[\sigma_{y1}\tilde{\rho}_{\mathrm{cq}}(t)\right]$
and $\langle\sigma_{i}\otimes\sigma_{i}\rangle=\Theta\mathrm{Tr}_{\mathrm{eq}}\left[\sigma_{i}\otimes\sigma_{i}\tilde{\rho}
_{\mathrm{cq}}(t)\right]$
for $i=x,y$. Moreover, based on Eq.~(\ref{eq:expectation}), we have
$\langle\sigma_{z1}\rangle=\mathrm{Tr}_{\mathrm{eq}}\left[\sigma_{z1}\tilde{\rho}_{\mathrm{cq}}(t)\right]$
and $\langle\sigma_{z}\otimes\sigma_{z}\rangle=\mathrm{Tr}_{\mathrm{eq}}
\left[\sigma_{z}\otimes\sigma_{z}\tilde{\rho}_{\mathrm{cq}}(t)\right]$.
Finally, one can obtain the density
matrix of the chromophore-qubit pair in the original basis under
the zeroth order approximation as
\begin{equation}
\rho_{\mathrm{cq}}(t)=M_{\Theta}\circ\tilde{\rho}_{\mathrm{cq}}(t),
\end{equation}
where $\circ$ denotes the Hadamard product and
\begin{equation}
M_{\Theta}=\left(\begin{array}{cccc}
1 & \Theta & 1 & \Theta\\
\Theta & 1 & \Theta & 1\\
1 & \Theta & 1 & \Theta\\
\Theta & 1 & \Theta & 1
\end{array}\right).
\end{equation}

%--------------------------Figure 9------------------------
\begin{figure}
\includegraphics[width=6.5cm]{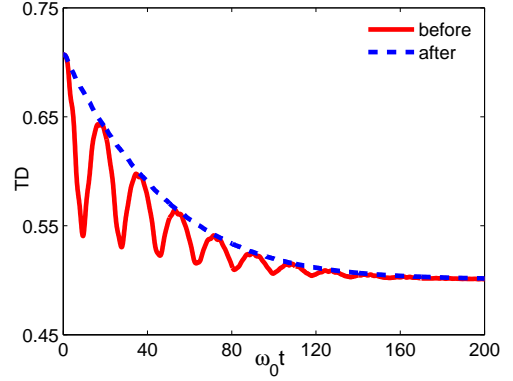}
\caption{\label{fig:TD_2qubit} Time evolution of the trace distance
for the chromophore-qubit pair. The parameters are
set as $\epsilon=\omega_{0}$, $\Delta=0.8\omega_{0}$, and $a=\omega_{0}$.
The spectral density are chosen as the same as that before. The red solid
line represents the trace distance before the polaron transformation
and the blue dashed line represents that after the transformation.}
\end{figure}
%--------------------------------------------------------------

Now we are in a position to study the non-Markovian behavior of the chromophore-qubit pair before the polaron transformation and that of the phonon-dressed quasi-particle.
Figure~\ref{fig:TD_2qubit} shows the time evolution of the trace
distances of the density matrices for the chromophore-qubit pair and the quasi-particle. The parameters in this plot are set as $\epsilon=\omega_{0}$, $\Delta=0.8\omega_{0}$, and $a=\omega_{0}$. The initial states of the chromophore-qubit pair are chosen
as $\rho_{\mathrm{c,half}}\otimes|0\rangle\langle0|_{\mathrm{q}}$
and $|0\rangle\langle0|_{\mathrm{c}}\otimes|0\rangle\langle0|_{\mathrm{q}}$.
Here $\rho_{\mathrm{c,half}}$ reads
\begin{equation}
\rho_{\mathrm{c,half}}=\frac{1}{2}\left(\begin{array}{cc}
1 & 1\\
1 & 1
\end{array}\right).
\end{equation}
After the polaron transformation, the initial states of the quasi-particle take the same form.
In Fig.~\ref{fig:TD_2qubit}, the solid and dashed lines depict the trace distance of
the density matrices before and after the polaron transformation,
respectively, and it is clear that before the polaron
transformation, there are oscillations in the trace-distance dynamics
implying non-Markovian behavior, while after
the polaron transformation, the trace distance monotonous decreases
with the passage of time, an observation in agreement with the well-known fact
that the polaron transformation reduces the effective interaction
between the quasi-particle and the bath.

\section{Conclusion} \label{sec:Conclusion}

In conclusion, utilizing a non-Markovian time-convolutionless
polaron master equation we probe dynamics of a central chromophore
interacting with a phonon reservoir via a probe qubit.
An in-depth analysis is carried out on the non-Markovian behavior
of the dynamics, for which a measure
of non-Markovianity is provided based on dynamical fixed points of the system.
This measure of non-Markovianity is analogous to the BLP measure but has a $\mathcal{O}(N)$
numerical advantage.

Using this measure, we have discussed the effects of the bath
temperature and the strength of the chromophore-qubit coupling on the non-Markovian behavior of the
chromophore. It is found that an increase in the temperature
brings about a reduction in the non-Markovianity.
In the weak coupling regime, an increase in the coupling is found to enhance the non-Markovianity,
while in the strong coupling regime, it suppresses the non-Markovianity.
The non-Markovianity maximum is found in the near resonance regime (around $a=\omega_{0}$) for
$T=0.5\omega_{0}$, a value that may be sensitive to other system parameters such as the temperature and the spectral density of the bath.
In addition,  we compare the non-Markovian behavior of the chromophore-qubit combination
before and after the polaron transformation. It is found that the non-Markovian
behavior vanishes after the polaron transformation.

Non-Markovianity is of great interest in a variety of topics, such as
steady-state entanglement maintenance~\cite{Huelge12}, quantum teleportation~\cite{Bylicaka13}, and precision estimation under noise in quantum metrology~\cite{Chin12}. Thus, finding optimal
conditions to maximize non-Markovianity is useful for many situations.
It is our hope that this work may inspire future experimental and theoretical endeavors to quantify non-Markovianity in relevant fields.

\begin{acknowledgments}
The authors thank Maxim Gerlin, Jian Ma, Dazhi Xu and
Lipeng Chen for useful discussion. One of us (JL) is especially indebted to
Xiao-Ming Lu for many valuable suggestions.
This work was supported by the Singapore National Research Foundation
through the Competitive Research Programme (CRP) under the Project
No.~NRF-CRP5-2009-04. One of us (XW) also acknowledges support from NFRPC
through Grant No.~2012CB921602 and NSFC through Grants Nos.~11025527
and 10935010.
\end{acknowledgments}

\end{document}